\documentclass{amsart}

\usepackage{graphicx}
\usepackage{eucal}

\newcommand{\one}{\ensuremath{\mathbf{1}}}
\newcommand{\sgn}{\operatorname{sgn}}
\newcommand{\rqtilde}{\ensuremath{\hspace{-1pt}_{\mathbf{\tilde{\;}}}}}

\newtheorem{theorem}{Theorem}
\newtheorem{prop}{Proposition}
\newtheorem{lemma}{Lemma}

\begin{document}

\title[Ubiquity of synonymity]{Ubiquity of synonymity: 
almost all large binary trees are not uniquely identified
by their spectra or their immanantal polynomials}

\author{Frederick A. \ Matsen}
\address{Frederick A. \ Matsen \\
Program for Evolutionary Dynamics \\
Harvard University \\
One Brattle Square, 6th Floor \\
Cambridge, MA 02138 \\
U.S.A.}
\email{matsen@math.harvard.edu}
\urladdr{http://www.math.harvard.edu/\rqtilde matsen/}

\author{Steven N.\ Evans}
\address{Steven N.\ Evans \\
  Department of Statistics \#3860 \\
  University of California at Berkeley \\
  367 Evans Hall \\
  Berkeley, CA 94720-3860 \\
  U.S.A.}
\email{evans@stat.Berkeley.EDU}
\urladdr{http://www.stat.berkeley.edu/users/evans/}
\thanks{SNE supported in part by NSF grant DMS-0405778 and part of the research was
conducted during a visit to the Pacific Institute for the Mathematical Sciences}

\keywords{tree statistic, algebraic graph theory, adjacency matrix, Laplacian matrix,
pairwise distance matrix, phylogeny, generating function,
functional equation}

\begin{abstract}
  There are several common ways to encode a tree as a matrix, such as
  the adjacency matrix, the Laplacian matrix (that is, the
  infinitesimal generator of the natural random walk), and the matrix
  of pairwise distances between leaves. Such representations involve a
  specific labeling of the vertices or at least the leaves, and so it
  is natural to attempt to identify trees by some feature of the
  associated matrices that is invariant under relabeling. An obvious
  candidate is the spectrum of eigenvalues (or, equivalently, the
  characteristic polynomial). We show for any of these choices of
  matrix that the fraction of binary trees with a unique spectrum goes
  to zero as the number of leaves goes to infinity. We investigate the
  rate of convergence of the above fraction to zero using numerical
  methods. For the adjacency and Laplacian matrices, we show that that
  the {\em a priori} more informative immanantal polynomials have no
  greater power to distinguish between trees.
\end{abstract}

\maketitle

\section{Introduction}

Tree shape theory furnishes numerical statistics about the structure of a
tree \cite{felsenstein, Mooers1997:31}.  (Because
we are interested in applications of tree statistics to
trees that describe evolutionary histories,
we will, for convenience, always take the term
{\em tree} without any qualifiers to mean a {\bf rooted, binary tree without any labeling of
the vertices}.)
Such statistics have two related uses.  Firstly, they can be used
in an attempt to tell whether two trees are actually the same
and, secondly, they can be used to indicate the degree of
similarity between  two trees with respect to some criterion.

Examples of the latter use are the testing of hypotheses about
macroevolutionary processes and the detection of bias in phylogenetic
reconstruction. Historically, numerical statistics for such purposes
have attempted to capture the notion of the {\em balance} of a tree,
which is the degree to which daughter subtrees are the same size. The
balance is typically measured by ad-hoc formulae that are often
selected for statistical power to distinguish between two different
distributions on trees \cite{Kirkpatrick1993:1171, Agapow2002:866}.

In this paper we take some steps in
investigating the possibility of a more mathematically 
``canonical'' algebraic approach to tree statistics
based on various matrix representations of the tree.  Our focus is
on the use of statistics for distinguishing trees rather than
quantifying degrees of similarity/dissimilarity.

We first describe the matrix representations of a tree that we will consider.

In algebraic graph theory \cite{MR1271140}, the basic matrix
associated to a graph is the {\em adjacency matrix} $A(G)$, whose
$ij^{\mathrm{th}}$ entry is one if $i$ and $j$ are connected by an
edge, and zero otherwise. From a probabilistic point of view, the more
natural matrix to associate with a graph is the {\em Laplacian matrix}
$L(G)$, which is the infinitesimal generator of the natural random
walk on the graph and is given by $D(G) - A(G)$, where $D(G)$ is the
diagonal matrix of vertex degrees. It is clear that a graph can be
recovered from either its adjacency of Laplacian matrix. Some authors,
such as \cite{MR1421568}, define the Laplacian to be $D(G)^{-1/2} L(G)
D(G)^{1/2}$. Note that this difference is not relevant if one is only
considering characteristics of the matrix $L$ such as eigenvalues that
are invariant under similarity transformations.

Readers familiar with the phylogenetics literature may be more
familiar with the {\em pairwise distance matrix} \cite{felsenstein, MR2060009}. 
The distance matrix $P$ given a leaf-labeling $1,
\ldots, n$ has as its $ij^{\mathrm{th}}$ entry the length of the path between
leaf $i$ and leaf $j$. Any leaf-labeled tree 
is uniquely determined by its distance matrix.   These
matrices have also been extensively studied as discrete metric spaces
\cite{MR1765236, matousek}. 

The definition of the adjacency and Laplacian matrices requires
a numbering of the vertices, while the definition of the
distance matrix requires a numbering of the leaves.
Because we are considering unlabeled trees (that is, we identify
trees that are equivalent in the usual sense of graph-theoretic isomorphism), we 
are only interested in tree statistics that are invariant under
renumbering.  Algebraically, this means that we are only interested
in features of the associated matrix that are unaffected by similarity transformations
via a permutation matrix.  The most obvious such statistics
are the eigenvalues.

The adjacency and Laplacian matrices and their eigenvalues are familiar objects in the area
of spectral graph theory \cite{MR1271140, MR1421568, MR1324340}. The
eigenvalues of the adjacency matrix tend to contain combinatorial
information about the graph, such as bounds on the chromatic number.
The eigenvalues of the Laplacian give information of a more geometric
flavor, such as the equivalent of the surface area to volume ratio of
subgraphs of a graph. As well as having
connections to the theory of random walks on graphs,
the Laplacian eigenvalues can be used
to define the expander graphs, an important class of graphs that have
applications in coding theory. Therefore, it would not be too
surprising if the these eigenvalues were a convenient way to summarize
information about a tree, giving a nice collection of tree
statistics.

Similarly, it seems plausible that the eigenvalues of the pairwise
distance matrix could contain quite a lot of information about the
tree that could be used to compare trees. Moreover, although the
distance matrix formally contains the same information as the
adjacency or Laplacian matrices, the transformation that takes the
distance matrix to one of the other two is distinctly non-linear, and
hence there is no reason to believe that there is any simple
connection between the corresponding eigenvalues.

We demonstrate below that not only do there exist pairs of
trees that have the same spectrum as another tree for the adjacency,
Laplacian, and distance matrices, but that this is the rule
rather than the exception as the trees become large, in the sense that
the fraction of trees with a given number of leaves
that have a unique adjacency or Laplacian spectrum goes to zero
as the number of leaves goes to infinity. 

The basic methodology that we use to prove this result was first
established in \cite{schwenk} and developed in \cite{MR1231010} for
general (that is, not necessarily bifurcating) graph-theoretic trees
in the case of the adjacency and Laplacian matrices. The present paper
provides the first results of this type concerning rooted bifurcating
trees, as well as the first examination of such results for the
pairwise distance matrix. The key idea is to establish that certain
pairs of trees $T_1$ and $T_2$ have the following {\em exchange
  property} for a given matrix representation: exchanging $T_1$ for
$T_2$ as subtrees of a given tree does not change the spectrum for
that matrix representation. It then becomes a matter of showing that
the number of trees with a given number of leaves is asymptotically of
larger order than the number of trees with the same number of leaves
that don't have a particular subtree. For this we build on the
generating function argument used in \cite{MR1502633} for asymptotic
estimates of the number of unlabeled rooted bifurcating trees (see
Section~\ref{sec:asymptotics}).

\begin{figure}
\label{fig:cospect}
\begin{center}
\includegraphics[angle=0,scale=0.8]{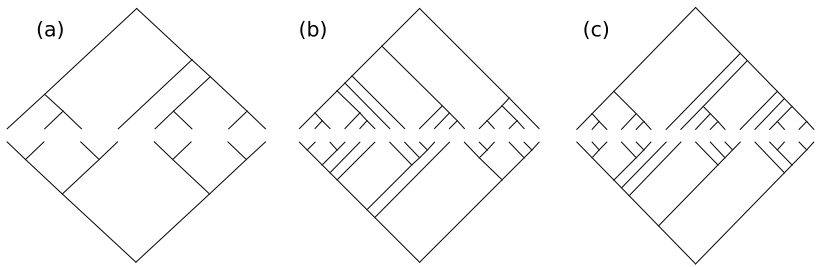}
\end{center}
\caption{Pairs of trees with similar algebraic properties. Figure (a)
  shows the smallest pair of rooted binary trees with the same
  adjacency and Laplacian spectrum. Figure (b) shows two trees with
  the exchange property for both the adjacency and Laplacian matrices.
  Figure (c) shows two trees with the exchange property for the
  pairwise distance matrix.}
\end{figure}

One possible explanation for this phenomenon is that two matrices
have the same spectrum if they are similar via an arbitrary similarity
transformation rather than just via a permutation transformation,
and this suggests considering features of a matrix 
that are invariant under permutation similarities
but not more general ones.  We will now describe a feature of
a matrix, its {\em immanantal polynomial}, that has this property.

The {\em immanant} is a generalization of the determinant. 
Recall that the determinant
of a matrix $A = \left(a_{ij}\right)$ is given by
\[
\mathrm{det} (A) := \sum_{\sigma \in S_n} \left. \sgn(\sigma) \prod_i a_{i \sigma(i)}
\right.,
\]
where the sum is over the symmetric group of permutations
of $\{1,2, \ldots, n\}$ and $\sgn(\sigma)$ is the {\em sign} of the permutation $\sigma$.

The function $\sgn$ is a particular example of a {\em character} of an
{\em irreducible representation} of the symmetric group.  It would take us
too far afield to define these notions here, but we note that
classical representation theory is one of the gems of pure
mathematics and excellent treatments
may be found in \cite{MR0125885, MR1153249, MR1363490, MR1824028}.  We note, however,
that the irreducible characters are constant on the conjugacy classes of
the symmetric group (recall that two permutations belong to the same
conjugacy class if and only if they have the same cycle structure) and they
form a basis for the vector space of functions with this property (the
{\em class functions}).

Our use of characters is simply to define the immanant 
\[
d_{\chi} (A) := \sum_{\sigma \in S_n} \left. \chi(\sigma) \prod_i a_{i \sigma(i)}
\right.
\]
of a matrix $A$ corresponding to the irreducible character $\chi$. A
discussion of immanants may be found in \cite{littlewood, MR0002127}.
The {\em immanantal polynomial} of a matrix is the corresponding
generalization of the characteristic polynomial; that is, it is the
polynomial $x \mapsto d_\chi(xI - A)$. Because the characters are
class functions, the immanantal polynomial is invariant under
similarity by permutation matrices, but it will not typically be
invariant under more general similarities.

Unfortunately, as we show in Lemma~\ref{lemma:spect_imm}, 
for either the adjacency
or Laplacian matrix the following two conditions on a pair
of trees are equivalent:
\begin{itemize}
\item
the spectra are equal,
\item
the immanantal polynomials  are equal for all
irreducible characters.
\end{itemize}
Consequently, the immanantal polynomials for the adjacency and
Laplacian matrices provide no more distinguishing power than the
spectra and, in particular, a vanishing fraction of large trees have a
unique immanantal polynomial for these matrices. We do not know if the
same fact is true for the distance matrix.

Our main result is thus the following.

\begin{theorem}
\label{theorem:main}
  Let $t_n$ be the number of trees with $n$ leaves.  For either the adjacency,
  Laplacian, or pairwise distance matrix, let 
  $l_n$ be the number of trees with $n$ leaves that do not share their spectrum
  with another tree. Then the fraction $l_n / t_n$ goes to zero as $n$ goes to
  infinity.  For the adjacency and Laplacian matrices, the same result holds 
  if we replace the spectrum by the complete set of immanantal polynomials.
\end{theorem}

The rate of convergence of the fraction in
Theorem~\ref{theorem:main} is also of interest. 
If it is extremely slow then 
the existence of trees with shared spectra may not be
practically relevant for the construction of informative tree shape statistics. We
investigate this matter numerically in Section \ref{sec:numer-exper}.

\section{Algebraic preliminaries concerning spectra and immanantal polynomials}

\subsection{Equality of adjacency and Laplacian spectra implies equality of immanantal polynomials}

In order to prove results for the adjacency and Laplacian matrices
simultaneously, 
we define for a tree $T$ and arbitrary real numbers $y$ and $z$
the {\em generalized Laplacian} $\tilde{L}(T) := yD(T) + zA(T)$  (recall that $A(T)$ is the adjacency
matrix and $D(T)$ is the diagonal matrix of vertex degrees).
We define the corresponding {\em generalized Laplacian immanantal
polynomial} of the tree $T$ with $r$ vertices to be
\[
x \mapsto d_{\chi} \left( xI - \tilde{L}(T) \right)
\]
for an irreducible character $\chi$ of the symmetric group $S_r$.

The generalized Laplacian immanantal polynomial can be computed in a simple combinatorial
fashion as follows. Define a {\em $k$-matching} to be a set of $k$ pairwise disjoint
edges of the tree (that is, a set of edges such that no two share
a common vertex). Let $M_k(T)$ denote the set of $k$-matchings on the
tree $T$.  We think of an edge as a pair of vertices,
so when we use the notation $i \notin p$ for a vertex
$i$ and an edge $p$ we mean that $i$ is not one of the
ends of $p$.   Let $C_k$ denote the conjugacy class of the symmetric group $S_r$
consisting of permutations that are the product of $k$ disjoint cycles,
and write $\chi(C_k)$ for the common value of the character $\chi$
on such permutations.  The following lemma appears in \cite{MR1231010} and is 
included for completeness.

\begin{lemma}
\label{lemma:match_poly}
The generalized Laplacian immanantal polynomial of the tree $T$
for the character $\chi$ is given by
\[
\sum_{k \geq 0} \chi(C_k) z^{2k} \sum_{p \in M_k(T)}
\prod_{i \notin p} (x - y d_i(T)).  
\]
\end{lemma}

\begin{proof}
Set $M := xI - \tilde{L}(T) = (m_{ij})$ so that the 
generalized Laplacian immanantal polynomial is
\begin{equation}
\label{eq:characteristic}
\sum_{\sigma \in S_n} \left. \chi(\sigma) \prod_i m_{i \sigma(i)}
\right..
\end{equation}
The matrix entries $m_{ij}$ are zero unless $i=j$ or there is an edge
between $i$ and $j$. If the permutation $\sigma$ has a cycle of
length $3$ or greater, then corresponding term in (\ref{eq:characteristic}) must
be zero because otherwise the tree would have a loop.
Therefore we need only consider permutations that are
products of disjoint transpositions
where, moreover, each transposition exchanges the two vertices of an edge. 
Such a permutation is equivalent in an obvious way to a $k$-matching
for some $k$, and the lemma follows.
\end{proof}

\begin{lemma}
\label{lemma:spect_imm}
Two trees have the same spectrum for their generalized Laplacian
if and only if they have the same generalized Laplacian immanantal polynomial
for all characters.
\end{lemma}

\begin{proof}
One direction is trivial: if two trees have the same 
generalized Laplacian immanantal polynomial
for all characters, then their generalized Laplacians have the 
same characteristic polynomial and hence the same spectrum.

Conversely, if the generalized Laplacians
of two trees have the same spectrum, then the characteristic
polynomials of the generalized Laplacians are the same.
Lemma~\ref{lemma:match_poly} in the case
$\chi = \sgn$, the fact that
$\sgn(C_k) = \pm 1$ for all $k$, and the fact that two equal
polynomials have the same coefficients imply that the
quantity
\[
\sum_{p \in M_k(T)} \prod_{i \notin p} (x - y d_i(T))
\]
is the same for both trees.  Another application of
Lemma~\ref{lemma:match_poly} completes
the proof.
\end{proof}

\subsection{A sufficient condition for two trees to have the same
adjacency or Laplacian spectrum}

We use the phylogenetic rather than graph-theoretic
definition of a subtree. That is, a subtree of a given rooted tree is
what results from separating an edge from its vertex furthest from
the root, which then
becomes the root of the subtree. 

Recall that $M_k(T)$
is the set of $k$-matchings of the tree $T$.
Let  $N_k(T)$ be the set of $k$-matchings where the chosen edges
do not contain the root. 

Define
\begin{eqnarray*}
P_k (T) & := & \sum_{p \in M_k(T)} \prod_{i \notin p} (x - y d_i(T)) \\
Q_k (T) & := & \sum_{p \in N_k(T)} \prod_{i \notin p} (x - y d_i(T)) 
\end{eqnarray*}

The following lemma is implicit in \cite{MR1231010}, but again we
include a proof for completeness.

\begin{lemma}
\label{lemma:matchings}
Let $S_1$ and $S_2$ be trees with the same number of leaves. 
If $P_k(S_1) = P_k(S_2)$ and 
$Q_k(S_1) = Q_k(S_2)$ for all $k$, then any tree with $S_1$ as a subtree has the same
generalized Laplacian spectrum as the tree obtained by substituting
$S_2$ for $S_1$. 
\end{lemma}

\begin{proof}
Let $T_1$ be a tree with $S_1$ as a subtree, and write $T_2$
for the tree obtained by substituting $S_2$ for $S_1$.
Denote by $e_0$ the edge that connects the rest of $T_1$ (resp. $T_2$)
to the root of $S_1$ (resp. $S_2$).

We differentiate between two types of $k$-matchings of $T_i$: those that
contain $e_0$ and those that do not. Note that a $k$-matching of $T_i$
that {\bf does not} contain $e_0$ restricts to an $\ell$-matching of $S_i$ for
some $\ell$, and all matchings of $S_i$ arise via such a restriction.  Similarly,
a $k$-matching of $T_i$ that {\bf does} contain $e_0$ restricts to an $\ell$-matching of $S_i$
with the property that the root of $S_i$ does not belong
to any edge in the matching, and all matchings of $S_i$ with this property arise 
via such a restriction.

Consider the formula for the characteristic polynomial of
the generalized Laplacian matrix that comes from
Lemma~\ref{lemma:match_poly} with $\chi=\sgn$.
Apply this formula to $T_1$ and $T_2$.
The assumption $P_k(S_1) = P_k(S_2)$ 
(resp. $Q_k(S_1) = Q_k(S_2)$) ensures that
the matchings that do not include (resp. do include) $e_0$ 
make the same contribution to the respective characteristic polynomials.
\end{proof}

The trees  depicted in Figure~\ref{fig:cospect}~(b) are
the smallest pair of rooted bifurcating trees satisfying the criteria
of this lemma. The verification of this fact was done by computer, and
the corresponding $P_k$ and $Q_k$ polynomials are available 
from the authors upon request.

\subsection{A sufficient condition for two trees to have the same
distance matrix spectrum}

We first recall an identity for determinants of partitioned matrices.
If
\[
C = 
\begin{pmatrix}
C_{11} & C_{12} \\
C_{21} & C_{22}
\end{pmatrix},
\]
then
\begin{equation}
\label{eq:det_id}
\begin{split}
\det C & = 
\det
\left(
\begin{pmatrix}
I & - C_{12} C_{22}^{-1} \\
0 & I
\end{pmatrix}
C
\begin{pmatrix}
I & 0 \\
- C_{22}^{-1} C_{21} & I
\end{pmatrix}
\right) \\
& = \det
\begin{pmatrix}
C_{11} - C_{12} C_{22}^{-1} C_{21} & 0 \\
0 & C_{22}
\end{pmatrix} \\
& = \det(C_{22}) \det(C_{11} - C_{12} C_{22}^{-1} C_{21}) \\
& = \det(C_{11}) \det(C_{22} - C_{21} C_{11}^{-1} C_{12}). \\
\end{split}
\end{equation}

\begin{lemma}
\label{lemma:distance_matrix}
Form two trees $T_1$ and $T_2$ by gluing
trees $S_1$ and $S_2$ with distance matrices $A_1$ and $A_2$ onto
the same leaf of a common tree $R$. Write $a_i$ for the vector of
distances from the leaves of $S_i$ to the root of $S_i$. 
Suppose that the following pairs of matrices have the same
spectra:
\[
A_i, \quad i=1,2,
\]
\[
\begin{pmatrix}
A_i & a_i \\
a_i' & 0
\end{pmatrix}, \quad i=1,2,
\]
\[
\begin{pmatrix}
A_i & a_i \\
\one' & 0
\end{pmatrix}, \quad i=1,2,
\]
\[
\begin{pmatrix}
A_i & \one \\
a_i' & 0
\end{pmatrix}, \quad i=1,2,
\]
and
\[
\begin{pmatrix}
A_i & \one \\
\one' & 0
\end{pmatrix}, \quad i=1,2,
\]
where $\one$ is a column vector with each entry $1$.
Then the distance matrices of $T_1$ and $T_2$ have the same spectrum.
\end{lemma}

\begin{proof}
Write $B$ for the distance matrix of $R$.  Then $B$ has the partitioned form
\[
\begin{pmatrix}
\check B & b \\
b' & 0
\end{pmatrix},
\]
where $\check B$ is the distance matrix of the tree obtained from $R$
by deleting the last leaf, $b$ is the column vector of distances from
the other leaves of $R$ to the last leaf. Assume without loss
of generality that this last leaf is the attachment point of the
$S_i$.

Denote by $D_i$ the distance matrix of $T_i$.  Observe that
\[
D_i=
\begin{pmatrix}
\check B & b \one' + \one a_i' \\
a_i \one' + \one b' & A_i
\end{pmatrix},
\]
Hence, by (\ref{eq:det_id}), $D_i$ has the characteristic polynomial
\[
\begin{split}
\det(xI - D_i)
& =
\det(xI - A_i) \det\bigl[(xI - \check B) - (-b \one' - \one a_i') (xI - A_i)^{-1} (-a_i \one' - \one b')\bigr] \\
& =
\det(xI - A_i) \det\Bigl[(xI - \check B) \\
& \quad -(\one'(xI - A_i)^{-1} a_i) \, b \one' \\
& \quad -(\one'(xI - A_i)^{-1} \one) \, b b' \\
& \quad -(a_i'(xI - A_i)^{-1} a_i) \, \one \one' \\
& \quad -(a_i'(xI - A_i)^{-1} \one) \, \one b' \Bigr].\\
\end{split}
\]

Using (\ref{eq:det_id}) again, we see that
a partitioned matrix of the form
\[
\begin{pmatrix}
A & g \\
h' & 0
\end{pmatrix},
\]
where $g$ and $h$ are column vectors,
has characteristic polynomial
\[
\det(xI - A) \bigl[x - h' (xI-A)^{-1} g \bigr],
\]
and the result follows.
\end{proof}

\section{Asymptotic numbers of trees}
\label{sec:asymptotics}

As outlined in the Introduction, the proof of 
Theorem~\ref{theorem:main} follows
immediately from Lemma~\ref{lemma:spect_imm},
Lemma~\ref{lemma:matchings}, Lemma~\ref{lemma:distance_matrix}, 
and the following result.

\begin{prop}
  \label{prop:asympt}
  Let \, $t_n$ be the number of trees with
  $n$ leaves. Let \, $s_n$ be the number of such trees that do not
  contain a given subtree. Then the fraction $s_n / t_n$ goes to zero
  as $n$ goes to infinity.
\end{prop}

\begin{proof}
Suppose that the forbidden subtree has $a$ leaves.
Let $f(x):= \sum_{i=1}^{\infty} t_i x^i$ and 
$f_a(x) := \sum_{i=1}^{\infty} s_i x^i$ denote the
generating functions for $t_n$ and $s_n$, respectively. 
Write $\rho$ for the radius of convergence of the power series
$f$ and $\rho_a$ for the radius of convergence of the power series
$f_a$.  Note that $\rho \le \rho_a < 1$.

It is shown in \cite{MR0025715} that
$\rho = 0.402698 \ldots$ and
\[
\lim_{n \rightarrow \infty} n^{3/2} \rho^n t_n = \eta,
\]
where $\eta = 0.7916032 \ldots$ (see \cite{MR0498168}
for an asymptotic expansion of $t_n$ that extends this
result and 
\cite{MR0406858, MR855396, MR0376369, MR0437344} 
for reviews of
general methods for determining asymptotic numbers of trees
of various sorts from a knowledge of the functional
equations that their generating functions solve).
Since $s_n$ is $o(\alpha^{-n})$ 
for any $\alpha < \rho_a$, it follows that 
$s_n/t_n$ is $o(\beta^n)$ for any $\beta > \rho/ \rho_a$,
and the proposition will hold 
if we can show that $\rho < \rho_a$.

For the sake of completeness and because it serves as
a good introduction to the derivation of the functional equation
satisfied by the generating function of $s_n$, we first derive
the well-known functional equation satisfied by the
generating function of the $t_n$.  See the comments after the
proof of the lemma for some remarks about the history of the
latter generating function. 

By decomposing a tree into the two subtrees rooted at the daughters
of the root, it is clear that
\begin{eqnarray*}
t_n = & t_1 t_{n-1} + t_2 t_{n-2} + \cdots + t_{m+1} t_{m-1}, &
\hbox{for} \ n = 2m+1, \\
t_n = & t_1 t_{n-1} + t_2 t_{n-2} + \cdots + t_{m+1} t_{m-1} + t_{m} (t_{m}+1)/2, &
\hbox{for} \ n = 2m.
\end{eqnarray*}
These expressions are equivalent to the statement 
\begin{equation}
\label{eq:recursion}
\sum_{i=1}^{n-1} t_i t_{n-i} = 2 t_n + t_{n/2}
\end{equation}
where $t_{n/2}$ is set to zero unless $n$ is even.

From (\ref{eq:recursion}) the generating function $f$ satisfies the functional equation
\begin{eqnarray*}
f^2(x) & = & \sum_{n=2}^{\infty} x^n \sum_{i=1}^{n-1} t_i t_{n-i} \\
& = & \sum_{n=2}^{\infty} x^n (2 t_n - t_{n/2}) \\
& = & 2 f(x) - f(x^2) - 2x.
\end{eqnarray*}
It will be convenient to consider the 
function $g := 1 - f$, which satisfies the functional equation
\begin{equation}
\label{eq:g}
g(x^2) = 2x + g^2(x).
\end{equation}
It is shown in \cite{MR1502633} that:
\begin{itemize}
\item
The radius of convergence $\rho$ is strictly positive.
\item
The functional equation (\ref{eq:g}) has a unique solution in the whole complex plane,
and this solution agrees with our power series in $\{x \in {\mathbb C}: |x| < \rho\}$.
\item
If, with a slight abuse of notation, we also denote this solution by $g$, then $g(\rho) = 0$.
\item
The point $\rho$ is the only zero of $g$ within $\{x \in {\mathbb C}: |x| < 1\}$.
\end{itemize}
It is clear from the power series that $g$ is continuous and decreasing on $[0,\rho)$ and $g(0) = 1$.
Hence $g$ is strictly positive on $[0,\rho)$.

As observed in \cite{MR1502633}, these observations suggest a method for computing $\rho$.
Put $h(x) =
g(x) / \sqrt{x}$, so that $h$ satisfies  $h(x^2) = 2 + h^2(x)$. 
Set 
\[
w_k(\eta) := \left(2 + \eta^{2^k}\right)^{2^{-k}}, \quad \eta \in {\mathbb R},
\]
and
\[
q_n := w_{n-1} \circ w_{n-2}  \cdots \circ w_0,
\]
so that each function $q_n$ is strictly increasing on $[-2, \infty)$ and $q_1 \le q_2 \le \ldots$.
In particular, $q_n$ converges pointwise as $n \rightarrow \infty$.
Moreover,
\[
\lim_{n \rightarrow \infty} q_n(h(x)) 
= \lim_{n \rightarrow \infty} (h(x^{2^n}))^{2^{1-n}} 
= \lim_{n \rightarrow \infty} \frac{(g(x^{2^n}))^{2^{1-n}}}{x} 
= \frac{1}{x}
\]
for $0 < x < 1$.  Therefore
\[
\frac{1}{\rho} = \lim_{n \rightarrow \infty} q_n(0).
\]

Conveniently,
(\ref{eq:recursion}) holds with $s_n$ in place of $t_n$ for all $n$ except
for $n=a$; in this case one simply adds two to the right hand side of
the equation to make up for the fact that $s_a = t_a - 1$. 
Hence $f_a$ satisfies the functional equation.
\begin{equation}
\label{eq:gf-deleted}
f_a^2(x) = 2 f_a(x) - f_a(x^2) - 2x + 2x^a.
\end{equation}

Set $g_a := 1 - f_a$, so that
\begin{equation}
\label{eq:ga}
g_a(x^2) = 2x - 2x^a + g_a^2(x).
\end{equation}
It is clear that $g_a$ is continuous and decreasing on $[0,\rho_a)$
and $g_a(0)=1$.  Following the arguments in \cite{MR1502633},
the functional equation (\ref{eq:ga}) has a unique solution
in the whole complex plane,
and this solution agrees with our power series in $\{x \in {\mathbb C}: |x| < \rho_a\}$.
Moreover, analogues of the other properties of $g$ obtained in \cite{MR1502633} hold for $g_a$.  

Set $h_a(x) = g_a(x) / \sqrt{x}$, so that
\[
h_a(x^2) = 2 - 2 x^{a-1} + h_a^2(x).
\]
Put 
\[
w_{k,a,\xi}(\eta) := \left(2 - 2 \xi^{2^k} + \eta^{2^k}\right)^{2^{-k}}, \quad \eta \in {\mathbb R},
\] 
and
\[
q_{n,a,\xi} := w_{n-1,a,\xi} \circ w_{n-2,a,\xi}  \cdots \circ w_{0,a,\xi}.
\]
Then
\[
\lim_{n \rightarrow \infty} q_{n,a,x^{a-1}}(h(x)) 
= \lim_{n \rightarrow \infty} (h_a(x^{2^n}))^{2^{1-n}} 
= \lim_{n \rightarrow \infty} \frac{(g_a(x^{2^n}))^{2^{1-n}}}{x} 
= \frac{1}{x}
\]
for $0 < x < 1$, and, in particular,
\[
\frac{1}{\rho_a} = \lim_{n \rightarrow \infty} q_{n,a,\rho_a^{a-1}}(0).
\]

Now 
\[
\begin{split}
q_{n,a,\rho_a^{a-1}}(0) & = w_{n-1,a,\rho_a^{a-1}} \circ w_{n-2,a,\rho_a^{a-1}}  \cdots \circ w_{0,a,\rho_a^{a-1}}(0) \\
& \le
w_{n-1} \circ w_{n-2}  \cdots \circ w_1 \circ w_{0,a,\rho_a^{a-1}}(0) \\
& =
q_n(-2\rho_a^{a-1}), \\
\end{split}
\]
and so
\[
\frac{1}{\rho_a} \le \lim_{n \rightarrow \infty} q_n(-2\rho_a^{a-1}) \le \lim_{n \rightarrow \infty} q_n(0) = \frac{1}{\rho}.
\]
It therefore suffices to show that the function $y \mapsto \lim_{n \rightarrow \infty} q_n(y)$ 
is strictly increasing on $(-\varepsilon,\infty)$ for some 
$0 < \varepsilon < 2$.

Observe that
\[
q_n' = \prod_{k=1}^{n-1} w_k' \circ q_{k}. 
\]
For $k \ge 1$,
\begin{eqnarray*}
w_k'(x) & = & x^{2^k-1} \left(2+x^{2^k}\right)^{2^{-k}-1} \\
& = & \left(2 x^{-2^k}+1\right)^{2^{-k} -1},
\end{eqnarray*}
so that $x \mapsto w_k'(x)$ is  non-decreasing for $x > 0$.
For $y \in (- \varepsilon, \infty)$,
\[
w_k'\circ q_k(y) \ge w_k' \circ q_1(y) \ge w_k' \circ q_1(-\varepsilon)
= w_k'(2 - \varepsilon)
\]
and hence
\[
\liminf_{n \rightarrow \infty} \inf_{y > -\varepsilon} q_n'(y) 
\ge 
\prod_{k=1}^{\infty} w_k'(2-\varepsilon).
\]

Taking $0 < \varepsilon < 1$,
the proof will be completed by demonstrating for any $x > 1$ that 
\[
\prod_{k=1}^{\infty} w_{k}' (x) > 0.
\]
Taking the logarithm gives
\begin{eqnarray*}
\sum_{k=1}^{\infty} (2^{-k} - 1) \, \log \left( 2 x^{-2^k} + 1 \right)
& > & - \sum_{k=1}^{\infty} \log \left( 2 x^{-2^k} + 1 \right), \\
& > & - \sum_{k=1}^{\infty} 2 x^{-2^k}, \\
\end{eqnarray*}
and this series clearly converges by the ratio test.
\end{proof}

In relation to Proposition~\ref{prop:asympt}, we note from
\cite{schwenk} that
the number of rooted strictly bifurcating trees without a given
subtree is asymptotically smaller than the number of all graph-theoretic
trees (see also \cite{MR1395455} for more about the enumeration of
general trees without a given subtree), but this is not enough for our purposes. 
We needed to show that it is
asymptotically smaller than the space of all rooted strictly
bifurcating trees.  The generating function for $t_n$ seems first to have
been investigated in \cite{MR1502633} 
in connection with enumerating ``types of arrangements'' in a commutative but
non-associative algebra, such as $a_1 (a_2 (a_3 a_4))$ or $(a_1 a_2)
(a_3 a_4)$; these are identical to rooted bifurcating trees in the
``Newick'' format \cite{felsenstein}.
The recursion behind the generating function has been
re-discovered independently several times such as in
\cite{Eth37} -- see \cite{MR1527424} for a discussion and several
further references.    We remark that
numerically iterating the quantity $q_n$
of the proof converges quickly to the value of
$\rho^{-1}$ calculable by other means \cite{MR0025715, MR0282451}.  We also
observe that
the methods of \cite{MR0498168, MR0406858, MR855396, MR0376369, MR0437344} can be used
to show, in the notation of the proof, that 
$\lim_{n \rightarrow \infty} n^{3/2} \rho_a^n s_n = \eta_a$ for some positive
constant $\eta_a$ and hence 
$\lim_{n \rightarrow \infty} (\rho_a/\rho)^n (s_n / t_n)  = \eta_a/\eta$,
but we don't pursue this matter here.

\section{Numerical experiments}
\label{sec:numer-exper}

Proposition~\ref{prop:asympt} says nothing about the rate of
convergence of the fraction. 
Here we investigate this rate using
computation. The characteristic polynomials for the generalized
Laplacian were calculated via a doubly-recursive algorithm to
enumerate matchings. The characteristic polynomials for the distance
matrices were calculated via the Leverrier-Faddeev algorithm
\cite{MR1642776}. These algorithms were implemented in \texttt{ocaml}
\cite{ocaml} and the code is available upon request.

\begin{table}
  \centering
  \begin{tabular}{c|ccc}
    leaves & trees & GLS & DS \\
    \hline
    2 & 1 & 1 & 1 \\
    3 & 1 & 1 & 1 \\
    4 & 2 & 2 & 2 \\
    5 & 3 & 3 & 3 \\
    6 & 6 & 6 & 6 \\
    7 & 11 & 11 & 11 \\
    8 & 23 & 22 & 23 \\
    9 & 46 & 45 & 46 \\
    10 & 98 & 95 & 98 \\
    11 & 207 & 203 & 207 \\
    12 & 451 & 443 & 451 \\
    13 & 983 & 972 & 983 \\
    14 & 2179 & 2159 & 2179 \\
    15 & 4850 & 4827 & 4850 \\
    16 & 10905 & 10870 & 10905 \\
    17 & 24631 & 24580 & 24630 \\
    18 & 56011 & 55931 & 56009 \\
  \end{tabular}
  \caption{The number of trees, the number of spectra for the
    generalized Laplacian, and the number of spectra for the distance
    matrix.}
  \label{tab:n-spect}
\end{table}

Table \ref{tab:n-spect} clearly shows that the fraction of trees with
unique spectra does not go to zero very quickly. Of course we can't
compute this fraction for large numbers of leaves, but we can get some
idea of the convergence by using the recursion relation corresponding
to the generating function (\ref{eq:gf-deleted}). We plot the
number of trees that do not have one of two subtrees of size
seventeen as a subtree. This is an actual fraction that 
can be used with
Proposition~\ref{prop:asympt} 
in order to prove Theorem~\ref{theorem:main}
for the generalized Laplacian matrix.

\begin{figure}
\label{fig:enumerate}
\begin{center}
\includegraphics[angle=0,scale=0.6]{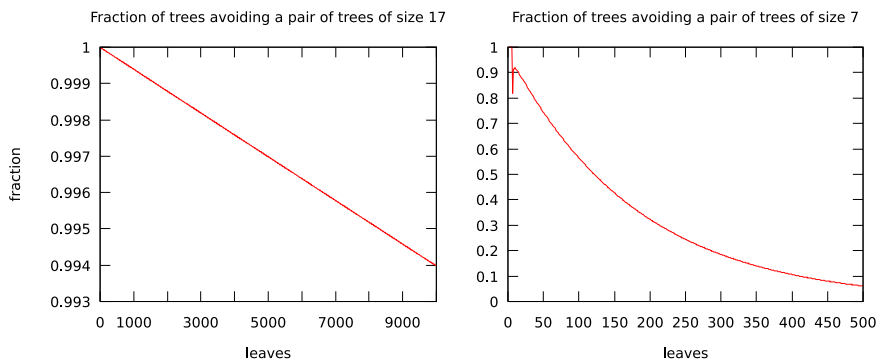}
\end{center}
\caption{The fraction of trees not containing a given pair of subtrees
  of size 7 and 17.}
\end{figure}

Figure 3 shows that this fraction converges
extremely slowly, despite the fact that as shown above it is 
asymptotically equivalent to $\beta^n$ for
some $0<\beta<1$. It is important to note, however, that this
fraction is probably
a very crude upper bound on the fraction of trees that share a
spectrum with another tree. To aid comparison we have included the
much more quickly converging number of trees without a pair of
subtrees of size seven. As can be seen in Table \ref{tab:n-spect}, the
actual number not sharing a spectrum goes down considerably more
quickly, though it is probably still the case that the vast majority
of trees of intermediate size should have their own spectra.

In conclusion, we have shown that the fraction of strictly bifurcating
rooted trees having a unique spectrum with respect to the adjacency,
Laplacian, or pairwise distance matrices goes to zero as the number of
leaves goes to infinity. This indicates that the spectra of these
matrices cannot perfectly distinguish between trees. However, because
the convergence to zero appears to be so slow, these spectra may still provide useful
information for tree sizes that appear in practice.  
We plan to investigate this matter further in a future article.

\bibliography{algebraic_rev3}
\bibliographystyle{amsalpha}

\end{document}